\documentclass[twoside]{dis08}
\usepackage[latin1]{inputenc}
\usepackage[dvips]{graphicx,epsfig,color}
\usepackage{wrapfig,rotating}
\usepackage{amssymb,amsmath,array}

\newcommand{\mysection}[1]{{\bf #1}. }

\begin{document}
\title{The SISCone and anti-$k_t$ jet algorithms}

\author{Gregory Soyez$^1$
\thanks{Work done under contract No. DE-AC02-98CH10886 with the US
  Department of Energy.}
%
%
\vspace{.3cm}\\
%
Brookhaven National Laboratory - Physics Department\\
Building 510, Upton, NY 11973 - USA
}

\maketitle

\begin{abstract}
  We illustrate how the midpoint and iterative cone (with progressive
  removal) algorithms fail to satisfy the fundamental requirements of
  infrared and collinear safety, causing divergences in the
  perturbative expansion. We introduce SISCone and the anti-$k_t$
  algorithms as respective replacements that do not have those
  failures without any cost at the experimental level.
\end{abstract}

\mysection{The general picture}
Jets are an important tool in hadronic physics and they will play a
predominant role at the LHC. By defining jets one aims at accessing,
from the final-state particles, the underlying hard parton-level
processes. Since a parton is not a well-defined object, a jet
definition is also not unique.

Two broad classes of jet definitions exist. The first one works by
defining a distance between pairs of particles, performing successive
recombinations of the pair of closest particles and stopping when all
resulting objects are too far apart. Algorithms within that class
differ by the definition of the distance, frequent choices
being $d_{ij}^2=\min(k_{t,i}^2,k_{t,j}^2)(\Delta
y_{ij}^2+\Delta\phi_{ij}^2)$ for the $k_t$ algorithm \cite{kt}, and
$d_{ij}^2=(\Delta y_{ij}^2+\Delta\phi_{ij}^2)$ for the
Cambridge-Aachen algorithm \cite{cam}.

Cone algorithms make up the second class, where jets are defined as
dominant directions of energy flow. One introduces the concept of {\em
  stable cone} as a circle of fixed radius $R$ in the $y-\phi$ plane
such that the sum of all the momenta of the particles within the cone
points in the same direction as the centre of the circle. Cone
algorithms attempt to identify all the stable cones. Most
implementations use a seeded approach to do so: starting from one seed
for the centre of the cone, one iterates until the cone is found
stable. The set of seeds can be taken as the set of initial particles
(sometimes over a $p_t$ threshold) or as the midpoints between
previously-found stable cones. As we shall see, this iterative method
fails to identify {\em all} the stable cones, leading to infrared or
collinear unsafety in the perturbative computations.

Cone algorithms can be split in two subclasses according to how they
deal with the fact that stable cones may overlap. Cone algorithms with
split--merge, identify the hardest overlapping pair of stable cones
and merge (split) them if they share more (less) than a fraction $f$
of the hardest cone. JetClu and midpoint are typical representatives
of that subclass. Cone algorithms with progressive removal start with
the hardest unclustered particle, iterate from there until a stable
cone is found and call it a jet. Its contents are removed and one
starts again with the remaining particles. The iterative cone is the
typical example of a subclass with the particular feature that hard
jets are fully conical.

The SNOWMASS accords have established a series of requirements that
any jet algorithm has to fulfil. These are basically that one can use
the algorithm for theoretical computations, {\em e.g.} it gives finite
perturbative results, as well as for experimental purposes, {\em e.g.}
it runs fast enough and has small corrections from hadronisation and
the underlying event.

We illustrate in these proceedings \cite{url} that midpoint and the
iterative cone fail to give finite perturbative results. We introduce
SISCone and the anti-$k_t$ algorithms as solutions to those problems
that do not spoil the experimental usability.

\begin{figure}[t]
\centering{\includegraphics[width=0.76\textwidth]{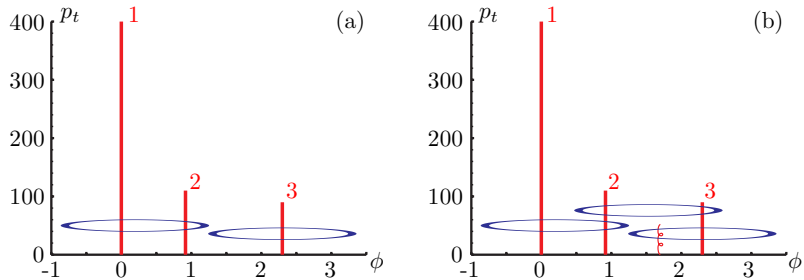}}
\caption{Stable cones found by the midpoint algorithm for a 3-particle
  event (left) and for the same event with an additional infinitely
  soft gluon (right).}\label{fig:irfailure}
\end{figure}

\mysection{SISCone as a replacement for the midpoint algorithm} Let us
consider the 3-particle event displayed in Fig.
\ref{fig:irfailure}(a). When clustered with the midpoint algorithm, 2
stable cones are found, leading to two jets: one with particles 1 and
2 and a second one with particle 3. If one adds to that hard event an
infinitely soft gluon as shown in Fig. \ref{fig:irfailure}(b), a third
stable cone is found and the three hard particles are clustered in a
single jet. This change in the jet structure upon addition of soft
particles, a phenomenon which happens with infinite probability in
perturbative QCD, gives rise to divergences in the perturbative
expansion and proves that the midpoint algorithm is infrared
unsafe.

This problem arises from the fact that the seeded approach misses
stable cones --- here the one containing particles 2 and 3 in
Fig. \ref{fig:irfailure}(a). The workaround to restore IR safety is
thus to find a seedless method that provably identifies all the stable
cones. This is notoriously complex: a naive approach testing the
stability of all subsets of particles \cite{blazey} has a complexity
of order $N2^N$ for $N$ particles which is much slower than the ${\cal
  O}(N^3)$ complexity of the midpoint algorithm, making this solution
unusable for experimental purposes.

\begin{wrapfigure}{r}{0.4\textwidth}
\includegraphics[width=0.4\textwidth]{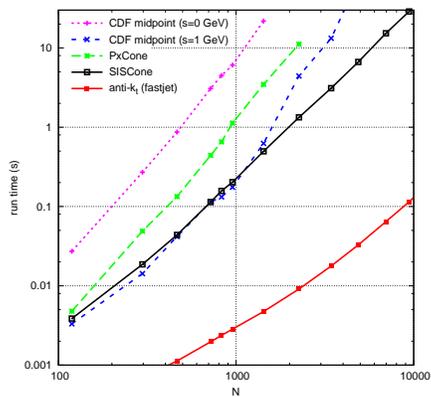}
\caption{Clustering time for SISCone compared to typical
  implementations of the midpoint algorithm and the anti-$k_t$
  algorithm \cite{fastjet}.}\label{fig:speed}
\end{wrapfigure}

The solution \cite{siscone} is to use the geometrical observation that
any enclosure in the $y-\phi$ plane can be moved without changing its
contents until it touches two points. Browsing all pairs of particles
allows thus to enumerate all possible cones and to check their
stability at an overall cost of ${\cal O}(N^3)$. Additional efforts
can even bring the final complexity to ${\cal O}(N^2\log(N))$ {\em
  i.e.}  faster than the midpoint algorithm. This is illustrated on
Fig. \ref{fig:speed} where we observe that in practice SISCone runs
faster than the typical implementations of the midpoint algorithm
without a seed threshold and at least as fast as when a 1 GeV seed
threshold is used.

This has been implemented \cite{siscone, siscone_code,fastjet} in
a \verb!C++!  code named SISCone (Seedless Infrared Safe Cone) which
is the first cone algorithm to satisfy the SNOWMASS requirements, that
is to be at the same time IR and collinear safe, and to be
fast enough to be used in experimental analysis.

\begin{figure}[t]
\centering{\includegraphics[width=0.76\textwidth]{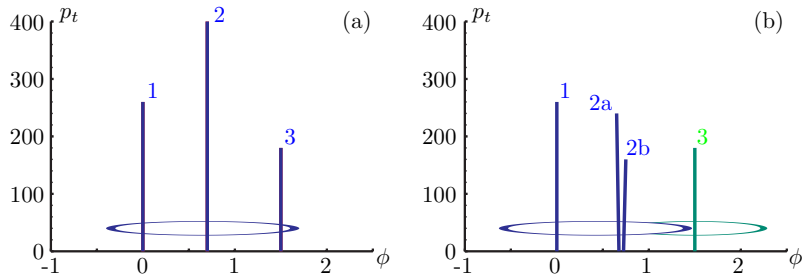}}
\caption{Jets found by the iterative cone for a 3-particle event
  (left) and for the same event with a collinear splitting
  (right).}\label{fig:uvfailure}
\end{figure}

\mysection{Anti-$\boldsymbol{k_t}$ as a replacement for the iterative cone}
As for the midpoint algorithm, we start with an event with three hard
particles (see Fig. \ref{fig:uvfailure}(a)). When clustered with the
iterative cone, one stable cone containing all particles is found,
resulting in a 1-jet event. If we now split the hardest particle into
two collinear particles --- a process that also has an infinite
probability in perturbative QCD --- as shown on
Fig. \ref{fig:uvfailure}(b), clustering with the iterative cone gives
a first jet made of particle 1 plus the two collinear ones, then a
second jet with particle 3. This example proves that the iterative
cone algorithm is collinear unsafe. 

Quite surprisingly, we can find a solution to that problem by coming
back to the class of the recombination algorithms. The distance
measures introduced earlier can be written as
\[
d_{ij}^2=\min(k_{t,i}^{2p},k_{t,j}^{2p})(\Delta y_{ij}^2+\Delta\phi_{ij}^2),
\]
with $p=1$ for the $k_t$ algorithm and $p=0$ for the Cambridge/Aachen
algorithm. We can then consider a third case, the one for which $p=-1$
and call it the {\em anti-$k_t$ algorithm} \cite{antikt}. Obviously,
this algorithm is IR and collinear safe. Furthermore, since its
implementation \cite{fastjet} is similar to the one of the $k_t$
algorithm, its speed will be similar too, which certainly makes it
usable for experimental purposes as seen on Fig. \ref{fig:speed}.

\begin{wrapfigure}{r}{0.48\textwidth}
\includegraphics[width=0.48\textwidth]{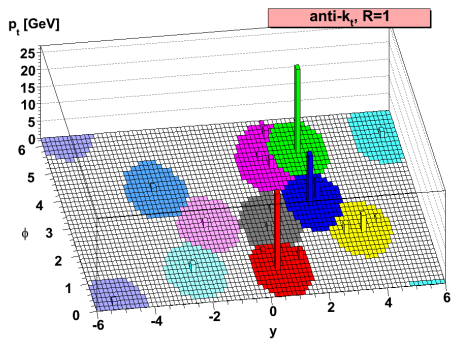}
\caption{Illustration of the regularity of the jets obtained with the
  anti-$k_t$ algorithm.}\label{fig:antikt}
\end{wrapfigure}

To understand the link with the iterative cone algorithm, we note from
the definition of the anti-$k_t$ distance that pairs involving a hard
particle will be given small distances. This means that soft
particles are recombined with hard ones before recombining among
themselves, resulting in regular, soft-resilient, hard jets. This is
exactly the hallmark of the iterative cone and it is in that respect
that the anti-$k_t$ can be seen as an IR and collinear safe
replacement.

To illustrate this property, we show in Fig. \ref{fig:antikt} the jets
resulting from the clustering of an event made with a few hard
particles and a large number of very soft ones uniformly
distributed. It is clear that the hardest jets are perfectly circular
and that, in general, the boundaries between the jets are regular.

\mysection{Physical impact and discussion}
As we have seen, the seeded approach to stable cone search suffers
from problems w.r.t. perturbative QCD expansion: the algorithms with
split--merge are IR unsafe, while the iterative cone (with progressive
removal) is collinear unsafe. We have introduced SISCone as a natural
replacement of the cone algorithms with split--merge like midpoint,
and the anti-$k_t$ algorithm as a candidate to replace the iterative
cone. These new algorithms are both IR and collinear safe.

The question one might ask is to what extend these IR and collinear
safety issue are important in real measurements. Since the unsafety
arises when one has 3 particles in a common vicinity, it becomes
important at the order $\alpha_s^4$ or $\alpha_{\text{EW}}\alpha_s^3$
of the perturbative series.

\begin{table}
\begin{center}
  \begin{tabular}{|l|c|c|}\hline
    Observable                     & 1st miss cones at & Last meaningful order \\ \hline
    Inclusive jet cross section    &       NNLO        &         NLO  \\
    $W/Z/H$ + 1 jet cross section  &       NNLO        &         NLO  \\
    $3$ jet       cross section    &        NLO        &  LO (NLO in NLOJet)   \\
    $W/Z/H$ + 2 jet cross sect.    &        NLO        &  LO (NLO in MCFM)   \\
    jet masses in $3$~jets         &         LO        &  {\bf none} (LO in NLOJet)\\
         \hline
  \end{tabular}
\end{center}
\caption{Perturbative level at which IR or collinear unsafety arises
  for various processes.}\label{tab:processes}
\end{table}

\begin{wrapfigure}{r}{0.4\textwidth}
\includegraphics[width=0.4\textwidth]{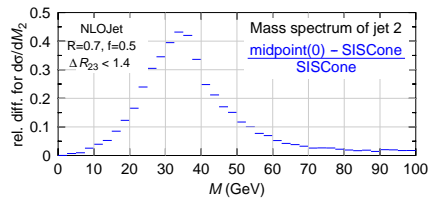}
\caption{Relative difference between midpoint and SISCone for the mass
  of the $2^{\rm nd}$ hardest jet in 3-jet events. The $2^{\rm nd}$ and
  $3^{\rm rd}$ jets are imposed to be distant by at most
  $2R$.}\label{fig:mass}
\end{wrapfigure}

Table \ref{tab:processes} summarises for different physical processes,
the order at which seeded algorithms stop to be valid. The main
message we can get from that table is thus that, if we do not want
theoretical efforts in precise QCD computations to be done in vain,
the resort of an IR and collinear safe algorithm like SISCone and the
anti-$k_t$ is fundamental.
To illustrate the argument more quantitatively, Fig. \ref{fig:mass}
shows the relative difference, expected to be present at the LO of
perturbative QCD, between SISCone and midpoint for the mass of the
$2^{\rm nd}$ hardest jet in 3-jet events. Differences reaching up to
40\% are observed, proving that an IR and collinear safe algotithm is
mandatory.
\vspace*{0.0cm}


\begin{footnotesize}


\end{footnotesize}


\end{document}